%% file: WoLLIC14main.tex
\documentclass[runningheads]{llncs}
\usepackage[T1]{fontenc}
\usepackage[all]{xy}
\usepackage{graphicx}
\usepackage{hyperref}
\usepackage{color}
\usepackage{mathtools}
\usepackage{amsmath}
\usepackage{amsfonts}
\usepackage{amssymb}
\usepackage{wasysym}
\usepackage{virginialake}
\usepackage{enumerate}
\usepackage{adjustbox}
\usepackage[shortlabels]{enumitem}

\usepackage{bussproofs}
\EnableBpAbbreviations
\input{macros}

\usepackage{thm-restate}
\begin{document}
\allowdisplaybreaks
\title{Paraconsistent Constructive Modal Logic}
%
%
\author{Han Gao\orcidID{0009-0004-1095-3347} \and
Daniil Kozhemiachenko\orcidID{0000-0002-1533-8034} \and
Nicola Olivetti\orcidID{0000-0001-6254-3754}}
\authorrunning{H.\ Gao et al.}
%
\institute{Aix-Marseille University, CNRS, LIS, Marseille, France\\
\email{\{gao.han,daniil.kozhemiachenko,nicola.olivetti\}@lis-lab.fr}}
\maketitle 

\begin{abstract}

We present a family of paraconsistent counterparts of the constructive modal logic $\ck$. These logics aim to formalise reasoning about contradictory but non-trivial propositional attitudes like beliefs or obligations. We define their Kripke-style semantics based on intuitionistic frames with two valuations which provide independent support for truth and falsity; they are connected by strong negation as defined in Nelson's logic. A family of systems is obtained depending on whether both modal operators are defined using the same or by different accessibility relations for their positive and negative support. We propose Hilbert-style axiomatisations for all logics determined by this semantic framework. We also propose a~family of modular cut-free sequent calculi that we use to establish decidability.

\keywords{constructive modal logic \and paraconsistent logic \and Nelson logic \and sequent calculi \and decidability.}
\end{abstract}

\input{01-introduction}
\input{02-logics}
\input{03-completeness}
\input{04-new-calculi-cut}
\input{05-cut}
\input{06-conclusion}
\bibliographystyle{splncs04}
\bibliography{mybibliography}
\input{appendix}

\end{document}

%% file: macros.tex
\newcommand{\hide}[1]{}

\newcommand{\Prop}{\mathsf{Prop}}
\newcommand{\simm}{{\sim}}
\newcommand{\Lnfourck}{\Lmc^\sim_{\Box,\lozenge}}
\newcommand{\Lnfour}{\Lmc^\sim}

\newcommand{\nfour}{\mathsf{N4}}
\newcommand{\Hnfour}{\mathcal{H}\mathsf{N4}}
\newcommand{\Gnfour}{\mathcal{G}\mathsf{N4}}
\newcommand{\ck}{\mathsf{CK}}
\newcommand{\nfourck}{\mathsf{CN4K}}
\newcommand{\Hnfourck}{\mathcal{H}\mathsf{CN4K}}
\newcommand{\Int}{\mathsf{Int}}
\newcommand{\ik}{\mathsf{IK}}

\newcommand{\nfckbd}{\nfourck^\curlyvee}
\newcommand{\nfckpm}{\nfourck^\pm}
\newcommand{\nfckj}{\nfourck^{\Join}}
\newcommand{\nfckone}{\nfourck^1}

\newcommand{\cnfck}{\mathcal{G}{\nfourck}}
\newcommand{\cnfckbd}{\mathcal{G}{\nfckbd}}
\newcommand{\cnfckpm}{\mathcal{G}{\nfckpm}}
\newcommand{\cnfckj}{\mathcal{G}{\nfckj}}
\newcommand{\cnfckone}{\mathcal{G}{\nfckone}}

\newcommand{\Cmc}{\mathcal{C}}
\newcommand{\Dmc}{\mathcal{D}}

\newcommand{\Gmc}{\mathcal{G}}
\newcommand{\Hmc}{\mathcal{H}}

\newcommand{\Lmc}{{\mathcal{L}}}

\newcommand{\Rmc}{{\mathcal{R}}}

\newcommand{\Ffrak}{\mathfrak{F}}

\newcommand{\Mfrak}{{\mathfrak{M}}}

\newcommand{\sfrak}{\mathfrak{s}}

\newcommand{\Kmbf}{{\mathbf{K}}}

\newcommand{\Rmbf}{{\mathbf{R}}}

\newcommand{\hmbf}{\mathbf{h}}
\newcommand{\rmbf}{\mathbf{r}}

\newcommand{\Cmsf}{{\mathsf{C}}}

\newcommand{\Lmsf}{{\mathsf{L}}}

\newcommand{\pspace}{\mathsf{PSpace}}

\newcommand{\nec}{\mathbf{nec}}

\newcommand{\Gammabox}{\Gamma^\Box}
\newcommand{\Gammadianeg}{\Gamma^\lozenge_\sim}

\newcommand{\Diamondpm}{\lozenge^\pm}
\newcommand{\Boxpmsim}{\Box^\pm_\sim}
\newcommand{\Boxbdsim}{\Box^\curlyvee_\simm}
\newcommand{\Diamondbd}{\lozenge^\curlyvee}
\newcommand{\Boxjoin}{\Box^\Join}
\newcommand{\Diamondjoin}{\lozenge^\Join}
\newcommand{\Boxonejoin}{\Box^{1,\Join}}
\newcommand{\Diamondonejoin}{\lozenge^{1,\Join}}
\newcommand{\Boxonesim}{\Box^1_\sim}

\newcommand\seq[2]{#1\Rightarrow #2}

\newcommand{\cut}{\mathsf{cut}}
\newcommand\diasimm[1]{{#1}^\lozenge_\simm}

%% file: 01-introduction.tex
\section{Introduction}

The aim of developing an intuitionistic or constructive notion of modalities is old: it goes back to Fitch~\cite{Fitch1950} and Prawitz~\cite{Prawitz1966}. Among the different proposals, the so-called Constructive modal logic based on the work by Wijesekera~\cite{Wijesekera1990} and then De Paiva et al.~\cite{AlechinaMendlerdePaivaRitter2001,BellindePaivaRitter2001} is motivated by applications to computer sciences, such as type-theoretic interpretation and Curry-Howard correspondence, but also by contextual reasoning~\cite{dePaiva2003,MendlerdePaiva2005}, where $\Box\phi$ and $\lozenge\phi$ are interpreted as ‘$\phi$ holds in all contexts’ and ‘$\phi$ holds in some contexts’, respectively. 
The constructive view is also relevant for the epistemic~\cite{Williamson1992}, doxastic, and deontic~\cite{DalmonteGrelloisOlivetti2022} reading of modalities, for instance in the latter case it is customary to distinguish the ‘strong’ permission that $\phi$, supported explicitly by some norm, from a~‘weak’ permission asserting only that $\neg\phi$ is not obligatory. 

Constructive modal logics are characterised by their simple proof-theoretic presentation in terms of ($\cut$-free) Gentzen sequent calculi (cf., e.g.,~\cite{Wijesekera1990,Simpson1994phd,DalmonteGrelloisOlivetti2021}). Their semantics can be specified\hide{, among the others,} in terms of bi-relational Kripke models containing both a~preorder, as in Kripke frames for propositional intuitionistic logic ($\Int$), and an accessibility relation. The main difference with other proposals such as $\ik$ by Fisher-Servi~\cite{FisherServi1980} and Simpson~\cite{Simpson1994phd} is that no properties are assumed to relate the preorder and the accessibility relation. Both Wijesekera's logic\hide{, which we name here WK,} and constructive modal logic $\ck$ satisfy some of the conditions stated by Simpson~\cite{Simpson1994phd}: they are conservative extensions of $\Int$, they have the disjunction property, and the two modalities are independent. Notice that the hereditary condition, necessary to ensure the conservativity over $\Int$ is built in by the forcing conditions, without the need for specific frame conditions (as, e.g., is done in $\ik$, recent $\mathbf{FIK}$~\cite{BalbianiGaoGencerOlivetti2024} and Do\v{s}en's $\mathbf{HK}_\Box$~\cite{Dosen1985}).
 
In this work, we aim to define a paraconsistent counterpart of $\ck$. The logic we are seeking must have the following features:

\begin{enumerate}[noitemsep,topsep=2pt]
\item it has the property of constructive falsity: if $\lnot(\phi\land\chi)$ is provable then either $\lnot\phi$ or $\lnot\chi$ is provable;
\item $\phi\land\lnot\phi$ does not entail every proposition (paraconsistency);
\item contradictions are not equivalent: $p\land\lnot p$ is not equivalent to $q\land\lnot q$.
\end{enumerate}
Intuitionistic logic with its standard negation $\lnot$ satisfies none of the three conditions. Our starting point for the interpretation of propositional connectives is the well-known logic $\nfour$ by Nelson~\cite{Nelson1949}. In this logic, intuitionistic negation and falsity are replaced by so-called strong negation $\sim$ that satisfies the properties above.

On the other hand, paraconsistent modal logics provide a~more intuitive doxastic and deontic interpretations of modalities. Classically (and intuitionistically), to account for contradictory beliefs or obligations, one may consider a~non-normal or non-regular logic. In the first case, $\Box(\phi\wedge\chi)$ is not equivalent to $\Box\phi\wedge\Box\chi$. In the second case, $\Box\phi\rightarrow\Box\chi$ does not follow from $\phi\rightarrow\chi$. Still, even in these cases, all contradictions are equivalent. Hence, if an agent believes in one contradiction, they believe in \emph{all contradictions}. If an agent has one conflicting obligation, then all obligations are contradictory. In addition to that, even in the presence of contradictory beliefs and obligations, one might want to utilise characteristic features of normal and regular modalities. Both options are possible when using paraconsistent logics.

The aim of this paper is to define a family of paraconsistent constructive modal logics of increasing strength, all of which can be considered $\nfour$-like counterparts of $\ck$ in a loose sense. We get several systems from the weakest to strongest according to the relation between the two modal operators $\Box$, $\lozenge$ and their strong negations: $\simm\Box$, $\simm\lozenge$. In the weakest system, there is no relation among the four. In the strongest one, $\Box$ and $\lozenge$ are reducible to one another via strong negation. 

The corresponding semantic picture is to consider $\nfour$-models having one or more accessibility relations for defining the modal operators. Namely, the models of the weakest system contain {\em four} independent accessibility relations (one for each modality and for their negations); the models of the strongest logic interpret all modalities using the same relation. 

We provide strongly complete Hilbert axiomatisations for all logics and construct cut-free sequent calculi which we use to establish the decidability of our systems. Due to the limited space, some proofs are put in the appendix.


%% file: 02-logics.tex
\section{Nelson's logic and its modal expansions\label{sec:nelson}}
We begin with the presentation of the propositional Nelson's logic ($\nfour$). Fix a~countable set $\Prop$ of propositional variables and define its language $\Lnfour$:
\begin{align*}
\Lnfour\ni\phi&\coloneqq p\in\Prop\mid\simm\phi\mid(\phi\wedge\phi)\mid(\phi\vee\phi)\mid(\phi\rightarrow\phi)
\end{align*}
The semantics of $\nfour$ is defined on intuitionistic frames \emph{with two independent valuations} corresponding to the \emph{support of truth} (positive support) and \emph{support of falsity} (negative support) of formulas in states.
\begin{definition}[Semantics of $\nfour$]\label{def:nfoursemantics}
An~\emph{$\nfour$ model} is a~tuple $\Mfrak=\langle W,\leq,v^+,v^-\rangle$ with $W\neq\varnothing$, $\leq$ being a~partial preorder on $W$, and $v^+,v^-:\Prop\rightarrow2^W$ s.t.\ $w\in v^+(p)$, $w\leq w'$ imply $w'\in v^+(p)$ and $w\in v^-(p)$, $w\leq w'$ imply $w'\in v^-(p)$.
The notions of \emph{positive} and \emph{negative support} of a~formula in a~state ($\Mfrak,w\Vdash^+\phi$ and $\Mfrak,w\Vdash^-\phi$, respectively) are defined as follows.
\begin{align*}
\Mfrak,w\Vdash^+p&\text{ iff }w\in v^+(p)\quad
\Mfrak,w\Vdash^-p\text{ iff }w\in v^-(p)\\
\Mfrak,w\Vdash^+\simm\phi&\text{ iff }\Mfrak,w\Vdash^-\phi\quad
\Mfrak,w\Vdash^-\simm\phi\text{ iff }\Mfrak,w\Vdash^+\phi\\
\Mfrak,w\!\Vdash^+\!\phi\!\wedge\!\chi&\text{ iff }\Mfrak,w\!\Vdash^+\!\phi\text{ and }\Mfrak,w\!\Vdash^+\!\chi\\\Mfrak,w\!\Vdash^-\!\phi\!\wedge\!\chi&\text{ iff }\Mfrak,w\!\Vdash^-\!\phi\text{ or }\Mfrak,w\!\Vdash^-\!\chi\\
\Mfrak,w\Vdash^+\phi\vee\chi&\text{ iff }\Mfrak,w\Vdash^+\phi\text{ or }\Mfrak,w\Vdash^+\chi\\\Mfrak,w\Vdash^-\phi\vee\chi&\text{ iff }\Mfrak,w\Vdash^-\phi\text{ and }\Mfrak,w\Vdash^-\chi\\
\Mfrak,w\Vdash^+\phi\rightarrow\chi&\text{ iff }\forall w'\geq w:\Mfrak,w'\Vdash^+\phi\Rightarrow\Mfrak,w'\Vdash^+\chi\\
\Mfrak,w\Vdash^-\phi\rightarrow\chi&\text{ iff }\Mfrak,w\Vdash^+\phi\text{ and }\Mfrak,w\Vdash^-\chi
\end{align*}

We say that $\phi\!\in\!\Lnfour$ is \emph{$\nfour$-valid} if $\Mfrak,w\!\Vdash^+\!\phi$ in each $\nfour$-model $\Mfrak$ and $w\in\Mfrak$.
\end{definition}

As one can see, the positive support conditions coincide with the semantics of Intuitionistic logic. The negative support conditions are dual to them via the \emph{classical} De Morgan laws. Thus, $\nfour$ is a~conservative extension of the positive Intuitionistic logic. Thus, it possesses disjunctive property and its dual \emph{constructive falsity property}: if $\simm(\phi\wedge\chi)$ is $\nfour$-valid, then $\simm\phi$ or $\simm\chi$ is $\nfour$-valid. Note, however, that there are other ways of treating the falsity condition of constructive implications~\cite{Wansing2008}. Still, they either lead to a~contradictory logic or define the negation of the implication via \emph{co-implication}. We choose the Nelsonian negation of $\rightarrow$ because it does not result in the language expansion.

The Hilbert-style axiomatisation of $\nfour$ can be obtained as follows.
\begin{definition}[$\Hnfour$ --- the Hilbert calculus for $\nfour$]\label{def:Hnfour}
$\Hnfour$ contains the following axiom schemes and rules ($\phi\leftrightarrow\chi$ stands for $(\phi\rightarrow\chi)\wedge(\chi\rightarrow\phi)$).
\begin{description}
\item[$\Int^+$:] Instantiations of the axioms for the \emph{positive} Intuitionistic logic in $\Lnfour$.
\begin{align*}
\simm\simm&\coloneqq\simm\simm\phi\leftrightarrow\phi&\mathbf{DeM}_\wedge&\coloneqq\simm(\phi\wedge\chi)\leftrightarrow(\simm\phi\vee\simm\chi)\\
\mathbf{DeM}_\vee&\coloneqq\simm(\phi\vee\chi)\leftrightarrow(\simm\phi\wedge\simm\chi)&\mathbf{DeM}_\rightarrow&\coloneqq\simm(\phi\rightarrow\chi)\leftrightarrow(\phi\wedge\simm\chi)\\
\mathbf{mp}&\coloneqq\dfrac{\phi\quad\phi\rightarrow\chi}{\chi}
\end{align*}
\end{description}
\end{definition}

Let us now consider $\nfourck$ --- a~modal expansion of $\nfour$. As expected, the language $\Lnfourck$ of $\nfourck$ expands $\Lnfour$ with two modalities: $\Box$ and $\lozenge$. Semantically, $\nfourck$ is defined on intuitionistic frames with additional relations used to compute the support of truth and falsity of modalities.
\begin{definition}[$\nfourck$ frames]\label{def:frames}
A~\emph{$\nfourck$ frame} is a~tuple of the following form: $\Ffrak=\langle W,\leq,R^+_\Box,R^-_\Box,R^+_\lozenge,R^-_\lozenge\rangle$ with $W\neq\varnothing$, $\leq$ being a~partial preorder on $W$, and $R^\bullet_\heartsuit$ binary relations on $W$ with $\bullet\in\{+,-\}$ and $\heartsuit\in\{\Box,\lozenge\}$.

A~frame is called \emph{$\pm$-birelational} if $R^+_\Box=R^+_\lozenge$ and $R^-_\Box=R^-_\lozenge$, \emph{$\curlyvee$-bi\-re\-la\-ti\-o\-nal} if $R^+_\Box=R^-_\Box$ and $R^+_\lozenge=R^-_\lozenge$, \emph{$\Join$-birelational} if $R^+_\Box=R^-_\lozenge$ and $R^+_\lozenge=R^-_\Box$, and \emph{monorelational} if all four relations coincide.

For a~relation $R$ on $W$ and $w\in W$, we set $R(w)=\{w'\mid wRw'\}$.
\end{definition}

In the definition above, we have five different classes of frames. In the most general case, $\Box$ and $\lozenge$ are interpreted with two independent relations corresponding to the support of truth and the support of falsity. Birelational frames are obtained by identifying two pairs of accessibility relations. Finally, in monorelational frames, all accessibility relations coincide. Note that it makes sense to use different accessibility relations to treat positive and negative support of modalities. Indeed, if one assumes \emph{proofs} (i.e., finding evidence \emph{in favour of something}) to be independent of \emph{refutations} (finding evidence \emph{against something}), it is also reasonable to assume that the states used to \emph{prove} a~modal statement are different from the states used to \emph{refute}~it.
\begin{definition}[$\nfourck$ models]\label{def:nfourckmodels}
A~\emph{$\nfourck$ model} is a~tuple $\Mfrak=\langle\Ffrak,v^+,v^-\rangle$ with $\Ffrak$ being a~$\nfourck$ frame and $v^+,v^-:\Prop\rightarrow2^W$ being such that
\begin{align*}
w\in v^+(p)\text{ and }w\leq w'&\Rightarrow w'\in v^+(p)&w\in v^-(p)\text{ and }w\leq w'&\Rightarrow w'\in v^-(p)
\end{align*}
The notions of \emph{positive} and \emph{negative support} for propositional connectives are the same as in Definition~\ref{def:nfoursemantics}. The semantics of modalities is as follows.
\begin{align*}
\Mfrak,w\Vdash^+\Box\phi&\text{ iff }\forall w'\geq w~\forall w''\in R^+_\Box(w'):\Mfrak,w''\Vdash^+\phi\\
\Mfrak,w\Vdash^-\Box\phi&\text{ iff }\forall w'\geq w~\exists w''\in R^-_\Box(w'):\Mfrak,w''\Vdash^-\phi\\
\Mfrak,w\Vdash^+\lozenge\phi&\text{ iff }\forall w'\geq w~\exists w''\in R^+_\lozenge(w'):\Mfrak,w''\Vdash^+\phi\\
\Mfrak,w\Vdash^-\lozenge\phi&\text{ iff }\forall w'\geq w~\forall w''\in R^-_\lozenge(w'):\Mfrak,w''\Vdash^-\phi
\end{align*}
\end{definition}

\begin{definition}\label{def:validityandentailment}
We say that $\Gamma\subseteq\Lnfourck$ \emph{entails $\chi$} ($\Gamma\models_{\nfourck}\chi$) if:
for every $\Mfrak=\langle\Ffrak,v^+,v^-\rangle$ and $w\in\Mfrak$ s.t.\ $\Mfrak,w\Vdash^+\phi$ for all $\phi\in\Gamma$, then $\Mfrak,w\Vdash^+\chi$.
If $\Gamma=\varnothing$, we write $\nfourck\models\chi$ and say that $\chi$ is \emph{valid}.
\end{definition}

We note briefly that $\Box$ and $\lozenge$ are dual to one another via $\simm$. Namely, $\simm\Box\simm$ behaves like~$\lozenge$ w.r.t.\ $R^-_\Box$ and $\simm\lozenge\simm$ like $\Box$ w.r.t.\ $R^-_\lozenge$.

In what follows, we use the following notation: $\nfourck$ stands for the logic of all $\nfourck$ frames, $\nfourck^*$ for the logic of all $*$-birelational frames with $*\in\{\pm,\curlyvee,\Join\}$, and $\nfourck^1$ for the logic of all monorelational frames. The notions of validity and entailment in these logics are obtained in the same way as in Definition~\ref{def:validityandentailment}.

It is important to note that apart from the number of relations in the frame, there are no other conditions on the accessibility relations themselves. This differentiates our logics from those studied in~\cite{OdintsovWansing2003,OdintsovWansing2008} and~\cite{Sherkhonov2008} as in both these cases, the accessibility relations defining modalities and the ordering of the states in the frame were connected via the confluence condition. This means that our logics are weaker than those of Odintsov, Wansing, and Sherkhonov. In particular, $\mathsf{DK}^{dd\sim}$ from~\cite{Sherkhonov2008} extends $\nfourck^\pm$ and $\mathcal{CALC}^\Cmsf$ from~\cite{OdintsovWansing2008} extends $\nfourck^1$. Still, as expected, all $\nfourck$ logics have persistence. 
\begin{proposition}\label{prop:persistence}
Let $\Mfrak$ be a~$\nfourck$ model $w\leq w'$, and $\phi,\chi\in\Lnfourck$. Then:
\begin{align*}
\text{if }\Mfrak,w\Vdash^+\phi,&\text{ then }\Mfrak,w'\Vdash^+\phi&\text{if }\Mfrak,w\Vdash^-\chi,&\text{ then }\Mfrak,w'\Vdash^-\chi
\end{align*}
\end{proposition}

We finish the section by observing that every $\Lnfourck$ theory is satisfiable. Consider a~model with one state $w$, all relations being equal to $\{\langle w,w\rangle\}$, and all variables both true and false at $w$. It is easy to check by the induction on formulas that every $\phi\in\Lmc$ is true (and false) at $w$.
\begin{restatable}{proposition}{trivialmodel}\label{prop:trivialmodel}
There is a~monorelational $\nfourck$ model $\Mfrak$ and s.t.\ $\Mfrak,w\Vdash^+\phi$ for every $\phi\in\Lnfourck$ and $w\in\Mfrak$.
\end{restatable}


%% file: 03-completeness.tex
\section{Hilbert calculi for $\nfourck$ and its extensions\label{sec:completeness}}
Let us now axiomatise Nelson's modal logics over the different classes of frames we introduced in Definition~\ref{def:frames}. To facilitate the presentation, we first introduce the following technical notion.
\begin{definition}\label{def:unionofcalculi}
Let $\Cmc$ be a~calculus and $\Rmbf$ a~set of rules and axioms. We use $\Cmsf\oplus\Rmbf$ to denote the calculus obtained by adding the rules and axioms in~$\Rmbf$ to~$\Cmc$.
\end{definition}

\begin{definition}[Hilbert calculi for $\nfourck$ and its extensions]\label{def:Hnfourck}
Consider the following axiom schemes and rules.
\begin{align*}
\top_\Box&\coloneq\Box(\phi\rightarrow\phi)&\top_\lozenge&\coloneq\simm\lozenge\simm(\phi\rightarrow\phi)\\
\wedge_\Box&\coloneq(\Box\phi\wedge\Box\chi)\rightarrow\Box(\phi\wedge\chi)&\wedge_\lozenge&\coloneq(\simm\lozenge\phi\wedge\simm\lozenge\chi)\rightarrow\simm\lozenge(\phi\vee\chi)\\
\pm_\Box&\coloneq\Box(\phi\rightarrow\chi)\rightarrow(\lozenge\phi\rightarrow\lozenge\chi)&\pm_\lozenge&\coloneq\simm\lozenge\simm(\simm\phi\!\rightarrow\!\simm\chi)\!\rightarrow\!(\simm\Box\phi\!\rightarrow\!\simm\Box\chi)\\
\curlyvee_\Box&\coloneq\Box(\phi\!\rightarrow\!\chi)\!\rightarrow\!(\simm\Box\simm\phi\!\rightarrow\!\simm\Box\simm\chi)&\curlyvee_\lozenge&\coloneq\simm\lozenge\simm(\phi\rightarrow\chi)\rightarrow(\lozenge\phi\rightarrow\lozenge\chi)\\
\Join_\Box&\coloneq\Box\phi\leftrightarrow\simm\lozenge\simm\phi&\Join_\lozenge&\coloneq\lozenge\phi\leftrightarrow\simm\Box\simm\phi\\
\rmbf_\Box&\coloneq\!\dfrac{\vdash\phi\rightarrow\chi}{\vdash\!\Box\phi\!\rightarrow\!\Box\chi}~\rmbf_\lozenge\!\coloneq\!\dfrac{\vdash\phi\rightarrow\chi}{\vdash\!\lozenge\phi\!\rightarrow\!\lozenge\chi}&
\rmbf^\sim_\Box&\coloneq\!\dfrac{\vdash\simm\phi\!\rightarrow\!\simm\chi}{\vdash\!\simm\Box\phi\!\rightarrow\!\simm\Box\chi}~\rmbf^\sim_\lozenge\!\coloneq\!\dfrac{\vdash\simm\phi\rightarrow\simm\chi}{\vdash\!\simm\lozenge\phi\!\rightarrow\!\simm\lozenge\chi}
\end{align*}

The calculi are as follows:
\begin{align*}
\Hnfourck&=\Hnfour\oplus\{\top_\Box,\wedge_\Box,\rmbf_\Box,\rmbf^\sim_\Box,\top_\lozenge,\wedge_\lozenge,\rmbf_\lozenge,\rmbf^\sim_\lozenge\}\\
\Hnfourck^\pm&=\Hnfourck\oplus\{\pm_\Box,\pm_\lozenge\}\quad
\Hnfourck^\curlyvee=\Hnfourck\oplus\{\curlyvee_\Box,\curlyvee_\lozenge\}\\
\Hnfourck^{\Join}&=\Hnfourck\oplus\{\Join_\Box,\Join_\lozenge\}\quad
\Hnfourck^1=\Hnfourck^\pm\!\oplus\!\Hnfourck^\curlyvee\!\oplus\!\Hnfourck^{\Join}
\end{align*}
An \emph{$\Hmc\Lmsf$-derivation of $\phi$ from $\Gamma$} ($\Gamma\!\vdash_{\Hmc\Lmsf}\!\phi$) is a~finite sequence of formulas $\phi_1$, \ldots, $\phi_n\!=\!\phi$ s.t.\ every $\phi_i$ is an instance of an axiom scheme, belongs to $\Gamma$, or obtained from previous formulas by an application of a~rule (with modal rules being applied to theorems only). If $\phi$ is derivable from $\Gamma\!=\!\varnothing$, we write $\Hmc\Lmsf\!\vdash\!\phi$ and say that $\phi$ is \emph{$\Hmc\Lmsf$-provable}.

In the text below, we let $\Lmsf\in\{\nfour,\nfourck,\nfourck^\curlyvee,\nfourck^\pm,\nfourck^{\Join},\nfourck^1\}$ and use $\Hmc\Lmsf$ to denote its Hilbert calculus.
\end{definition}



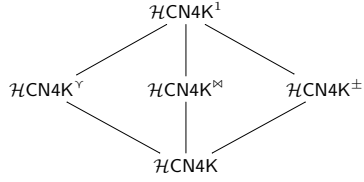
\begin{figure}[!t]
\centering
\begin{align*}
\resizebox{.4\linewidth}{!}{\xymatrix{&\Hnfourck^1\ar@{-}[d]\ar@{-}[dr]\ar@{-}[dl]&\\\Hnfourck^\curlyvee&\Hnfourck^{\Join}&\Hnfourck^\pm\\&\Hnfourck\ar@{-}[u]\ar@{-}[ur]\ar@{-}[ul]&}}
\end{align*}
\caption{Hilbert calculi for $\nfourck$ and its extensions.
}
\label{fig:hilbertcalculilattice}
\end{figure}

Observe that $\top_\heartsuit$, $\wedge_\heartsuit$, and $\rmbf_\heartsuit$ ($\heartsuit\in\{\Box,\lozenge\}$) can be replaced with the following axioms and rules.
\begin{align*}
\Kmbf_\Box&\coloneq\Box(\phi\rightarrow\chi)\rightarrow(\Box\phi\rightarrow\Box\chi)&\Kmbf_\lozenge&\coloneq\simm\lozenge\simm(\phi\rightarrow\chi)\rightarrow(\simm\lozenge\simm\phi\!\rightarrow\!\simm\lozenge\simm\chi)\\
\nec_\Box&\coloneq\dfrac{\vdash\phi}{\vdash\Box\phi}&\nec_\lozenge&\coloneq\dfrac{\vdash\phi}{\vdash\simm\lozenge\simm\phi}
\end{align*}

Furthermore, as one can see from Definition~\ref{def:nfourckmodels} \cite{BellindePaivaRitter2001}, and~\cite{DalmonteGrelloisOlivetti2021}, the positive support conditions of all connectives except $\sim$ coincide with the semantics of $\ck$. Thus, \emph{positive (i.e., $\sim$-free) mono-modal fragments} of $\nfourck$ coincide with the positive ($\bot$-free) mono-modal fragments of $\ck$. Moreover, the positive fragments of $\nfourck^\pm$ and $\nfourck^1$ coincide with the positive fragment of $\ck$.

The soundness can be obtained by checking the validity of axioms and that the rules preserve truth.
\begin{theorem}[Soundness]\label{theorem:soundness}
Let $\Gamma\vdash_{\Hmc\Lmsf}\phi$. Then $\Gamma\models_{\Lmsf}\phi$.
\end{theorem}

For the completeness, we follow the method of~\cite{Wijesekera1990}. We begin with an observation and some technical notions.
\begin{lemma}\label{lemma:modalitydistribution}
$\Hnfourck\vdash\bigwedge^{n}_{i\!=\!1}\!\!\Box\phi_i\!\leftrightarrow\!\Box\bigwedge^{n}_{i\!=\!1}\!\!\phi_i$ and $\Hnfourck\vdash\bigwedge^{n}_{i=1}\!\!\simm\lozenge\phi_i\!\leftrightarrow\!\simm\lozenge\!\bigvee^{n}_{i\!=\!1}\!\phi_i$.
\end{lemma}
\begin{definition}\label{def:saturatedsets}
We call $\Xi\subseteq\Lmc$ \emph{saturated} if it is \emph{deductively closed} (i.e., $\xi\in\Xi$ for $\Xi\vdash_{\Hnfourck}\xi$) and \emph{prime} (i.e., $\xi\vee\xi'\in\Xi$ entails that $\xi\in\Xi$ or $\xi'\in\Xi$).
\end{definition}

The next statement is proved by a~standard argument for the Lindenbaum lemma. Note that we do not need the usual requirement of \emph{consistency} since every $\Xi\subseteq\Lmc$ has a~model (recall Proposition~\ref{prop:trivialmodel}).
\begin{proposition}\label{prop:saturatednoderivation}
If $\Gamma\!\nvdash_{\Hnfourck}\!\phi$, there is a~saturated set $\Gamma'\!\supseteq\!\Gamma$ s.t.\ $\Gamma'\!\nvdash_{\Hnfourck}\!\phi$.
\end{proposition}

\begin{definition}\label{def:segments}
A~\emph{segment} is a~tuple $\sfrak=\langle\Xi,\Phi^+_\Box,\Phi^-_\Box,\Phi^+_\lozenge,\Phi^-_\lozenge\rangle$ s.t.\ $\Xi$ is a~saturated set that we call \emph{head} of $\sfrak$ ($\hmbf(\sfrak)$) and $\Phi^\bullet_\heartsuit$'s are families of saturated sets s.t.\ the following holds:
\begin{itemize}[noitemsep,topsep=2pt]
\item if $\Box\phi\in\Xi$, then $\phi\in\Delta$ for every $\Delta\in\Phi^+_\Box$;
\item if $\simm\Box\phi\in\Xi$, then $\simm\phi\in\Delta$ for some $\Delta\in\Phi^-_\Box$;
\item if $\lozenge\phi\in\Xi$, then $\phi\in\Delta$ for some $\Delta\in\Phi^+_\lozenge$;
\item if $\simm\lozenge\phi\in\Xi$, then $\simm\phi\in\Delta$ for every $\Delta\in\Phi^-_\lozenge$.
\end{itemize}
\end{definition}
\begin{definition}\label{def:canonicalmodel}
A~\emph{canonical $\nfourck$-model} is the following tuple: $\Mfrak^\Cmsf=\langle W^\Cmsf,\leq^\Cmsf,{R^+_\Box}^\Cmsf,{R^-_\Box}^\Cmsf,{R^+_\lozenge}^\Cmsf,{R^-_\lozenge}^\Cmsf,{v^+}^\Cmsf,{v^-}^\Cmsf\rangle$ s.t.\ $W^\Cmsf$ is the set of all segments and for any segments $\sfrak=\langle\Xi,\Phi^+_\Box,\Phi^-_\Box,\Phi^+_\lozenge,\Phi^-_\lozenge\rangle$ and $\sfrak'=\langle\Xi',{\Phi^+_\Box}',{\Phi^-_\Box}',{\Phi^+_\lozenge}',{\Phi^-_\lozenge}'\rangle$, we have:
\begin{align*}
\sfrak\leq^\Cmsf\sfrak'&\text{ iff }\hmbf(\sfrak)\subseteq\hmbf(\sfrak')&\sfrak{R^\bullet_\heartsuit}^\Cmsf\sfrak'&\text{ iff }\hmbf(\sfrak')\in\Phi^\bullet_\heartsuit\\
\Mfrak^\Cmsf,\sfrak\Vdash^+p&\text{ iff }p\in\hmbf(\sfrak)&\Mfrak^\Cmsf,\sfrak\Vdash^-p&\text{ iff }\simm p\in\hmbf(\sfrak)
\end{align*}
\end{definition}
\begin{restatable}[Truth lemma for $\Hnfourck$]{lemma}{truthlemmabasic}\label{lemma:truthlemmabasic}
For every $\phi\in\Lnfourck$ and $\sfrak\in\Mfrak^\Cmsf$, it holds that (1) $\Mfrak^\Cmsf,\sfrak\Vdash^+\phi$ iff $\phi\in\hmbf(\sfrak)$; (2) $\Mfrak^\Cmsf,\sfrak\Vdash^-\phi$ iff $\simm\phi\in\hmbf(\sfrak)$.
\end{restatable}
\begin{proof}
We proceed by induction on $\phi\in\Lnfourck$. 
If $\phi=p$, then the statement holds by the construction of $\Mfrak^\Cmsf$. For the induction steps, we only consider the case $\phi=\simm\lozenge\chi$. We assume that $\sfrak=\langle\Xi,\Phi^+_\Box,\Phi^-_\Box,\Phi^+_\lozenge,\Phi^-_\lozenge\rangle$.

Let $\phi=\simm\lozenge\chi$ and assume for contradiction that $\Mfrak^\Cmsf,\sfrak\Vdash^-\lozenge\chi$ but $\simm\lozenge\chi\notin\hmbf(\sfrak)$. Consider $\Xi^{\sim!}=\{\simm\tau\mid\simm\lozenge\tau\in\hmbf(\sfrak)\}$. Let us show that $\Xi^{\sim!}\nvdash_{\Hnfourck}\simm\chi$. Suppose for contradiction that there is some finite $\Xi'\subseteq\Xi^{\sim!}$ s.t.\ $\Xi'\vdash_{\Hnfourck}\simm\chi$. Then applying the deduction theorem, De Morgan laws for $\wedge$ and $\vee$, and $\rmbf^\sim_\lozenge$, we have $\Hnfourck\vdash\simm\lozenge\!\!\!\!\bigvee\limits_{\simm\tau'\in\Xi'}\!\!\!\!\tau'\rightarrow\simm\lozenge\chi$. Now, we use Lemma~\ref{lemma:modalitydistribution} which gives us $\Hnfourck\vdash\bigwedge\limits_{\simm\tau'\in\Xi'}\!\!\!\!\!\!\simm\lozenge\tau'\rightarrow\simm\lozenge\chi$. But as $\simm\lozenge\tau$'s belong to $\hmbf(\sfrak)$ which is deductively closed, this means that $\simm\lozenge\chi\in\hmbf(\sfrak)$ as well which contradicts our assumption. Thus, we can extend $\Xi^{\sim!}$ to a~saturated $\Xi^{\sim\sharp}$ s.t.\ $\Xi^{\sim\sharp}\nvdash_{\Hnfourck}\simm\chi$ and consider $\sfrak'=\langle\Xi,\Phi^+_\Box,\Phi^-_\Box,\Phi^+_\lozenge,\{\Xi^{\sim\sharp}\}\rangle$. Now let $\sfrak''$ be a~segment with $\hmbf(\sfrak'')=\Xi^{\sim\sharp}$. It is easy to see that $\sfrak'{R^-_\lozenge}^\Cmsf\sfrak''$ and that $\Mfrak^\Cmsf,\sfrak''\nVdash^-\chi$ by the induction hypothesis. Hence, using that $\sfrak\leq^\Cmsf\sfrak'$, we obtain $\Mfrak^\Cmsf,\sfrak\nVdash^-\lozenge\chi$, contrary to the assumption.

Conversely, let $\simm\lozenge\chi\in\hmbf(\sfrak)$. We have that $\simm\lozenge\chi\in\hmbf(\sfrak')$ for every $\sfrak'\geq^\Cmsf\sfrak$. Hence, $\simm\chi\in\hmbf(\sfrak'')$ for every $\sfrak''\in{R^-_\lozenge}^\Cmsf(\sfrak')$. Applying the induction hypothesis, we have $\Mfrak^\Cmsf,\sfrak''\Vdash^-\chi$ for every such $\sfrak''$ as required.
\qed
\end{proof}

We can now adapt the proof of Lemma~\ref{lemma:truthlemmabasic} to obtain the completeness of extensions of $\Hnfourck$. To do that, we introduce additional types of segments.
\begin{definition}\label{def:othersegments}
We say that $\sfrak$ is a~\emph{$\pm$-segment} if $\Phi^+_\Box=\Phi^+_\lozenge$ and $\Phi^-_\Box=\Phi^-_\lozenge$; \emph{$\curlyvee$-segment} if $\Phi^+_\Box=\Phi_\Box^-$ and $\Phi^+_\lozenge=\Phi^-_\lozenge$; \emph{${\Join}$-segment} if $\Phi^+_\Box=\Phi^-_\lozenge$ and $\Phi^+_\lozenge=\Phi^-_\Box$; \emph{$1$-segment} if $\Phi^+_\Box=\Phi^-_\lozenge=\Phi^+_\lozenge=\Phi^-_\Box$.

We let $*\in\{\pm,\curlyvee,\Join,1\}$. The canonical models $\Mfrak^\Cmsf_*$ of $\nfourck^*$'s can be defined as before (cf.~Definition~\ref{def:canonicalmodel}) but over the corresponding set of $*$-segments.
\end{definition}

\begin{restatable}{lemma}{truthlemmaextensions}\label{lemma:truthlemmaextensions}
It holds that (1) $\Mfrak^\Cmsf_*,\sfrak\!\Vdash^+\!\phi$ iff $\phi\!\in\!\hmbf(\sfrak)$, and (2) $\Mfrak^\Cmsf_*,\sfrak\!\Vdash^-\!\phi$ iff $\simm\phi\!\in\!\hmbf(\sfrak)$, for every $\phi\in\Lnfourck$ and $\sfrak\in\Mfrak^\Cmsf_*$.
\end{restatable}

The following statement now follows from Theorem~\ref{theorem:soundness} and Lemmas~\ref{lemma:truthlemmabasic}, and~\ref{lemma:truthlemmaextensions}.
\begin{restatable}{theorem}{Hnfourckcompleteness}\label{theorem:Hnfourckcompleteness}
$\Gamma\models_\Lmsf\phi$ iff $\Gamma\vdash_{\Hmc\Lmsf}\phi$.
\end{restatable}

We finish the section by noting an important asymmetry between the pairs of axioms extending $\Hnfourck$. While one could expect that adding any two pairs of axioms $\pm$, $\curlyvee$, and $\Join$ proves the remaining pair, this is not always the case.
\begin{theorem}\label{prop:incompleteness}~
\begin{enumerate}[noitemsep,topsep=2pt]
\item $\Hnfourck^\pm\oplus\{\Join_\Box,\Join_\lozenge\}\vdash\curlyvee_\Box\wedge\curlyvee_\lozenge$ and $\Hnfourck^\curlyvee\oplus\{\Join_\Box,\Join_\lozenge\}\vdash\pm_\Box\wedge\pm_\lozenge$.
\item $\Hnfourck^\pm\oplus\{\curlyvee_\Box,\curlyvee_\lozenge\}\nvdash\Join_\Box$ and $\Hnfourck^\pm\oplus\{\curlyvee_\Box,\curlyvee_\lozenge\}\nvdash\Join_\lozenge$.
\end{enumerate}
\end{theorem}
\begin{proof}
Item~1 is straightforward since $\Join$-axioms establish the interdefinability of $\Box$ and $\lozenge$. We tackle Item~2. 
Consider the following two frames over $W=\{w_0,w_1\}$ s.t.\ $w_0\leq w_1$:
\begin{align*}
\Ffrak_1&=\!\langle W,\leq,R^+_\Box,R^-_\Box,R^+_\lozenge,R^-_\lozenge\rangle: R^+_\lozenge=R^-_\Box\!=\!\varnothing, R^+_\Box\!=\!\{(w_0,w_0)\}, R^-_\lozenge\!=\!\{(w_0,w_1)\}\\
\Ffrak_2&=\!\langle W,\leq,S^+_\Box,S^-_\Box,S^+_\lozenge,S^-_\lozenge\rangle:S^+_\Box\!=\!S^-_\Box\!=\!S^-_\lozenge\!=\!\{(w_0,w_1),(w_1,w_1)\},S^+_\lozenge\!=\!\varnothing
\end{align*}

We show that $\Ffrak_1\models_\nfourck\Hnfourck^\pm\oplus\Hnfourck^\curlyvee$. Indeed, since both $R^+_\lozenge$ and $R^-_\Box$ are empty, $\pm$'s and $\curlyvee$'s axioms are vacuously valid (and other axioms and rules are valid on every frame). Now let $\Mfrak_1$ be a~model on $\Ffrak_1$ s.t.\ $\Mfrak_1,w_0\Vdash^+p$ and $\Mfrak_1,w_1\nVdash^+p$. Clearly, $\Mfrak_1,w_0\Vdash^+\Box p$. On the other hand, $\Mfrak_1,w_0\nVdash^-\lozenge\simm p$ (whence, $\Mfrak_1,w_0\nVdash^+\simm\lozenge\simm p$). Thus, $\Ffrak_1\not\models_\nfourck\Box p\leftrightarrow\simm\lozenge\simm p$.

Similarly, $\Ffrak_2\models_\nfourck\curlyvee_\lozenge$ because $S^+_\lozenge=\varnothing$. The validity of other axioms is also straightforward to establish. Now, assume $\Mfrak_2$ to be a~model on $\Ffrak_2$ s.t.\ $\Mfrak_2,w_1\Vdash^+p$. It is clear that $\Mfrak_2,w_1\Vdash^-\Box\simm p$ (i.e., $\Mfrak_2,w_1\Vdash^+\simm\Box\simm p$) but $\Mfrak_2,w_1\nVdash^+\lozenge p$ (as $S^+_\lozenge=\varnothing$). Thus, $\Ffrak_2\not\models_\nfourck\lozenge p\leftrightarrow\simm\Box\simm p$.
\qed
\end{proof}

%% file: 04-new-calculi-cut.tex
\section{Sequent calculi\label{sec:sequents}}

In this section, we present sequent calculi for all logics under consideration. Below $\Gmc\Lmsf$ denotes the sequent calculus of~$\Lmsf$.


\begin{figure}[!t]
\centering
\begin{center}
$\mathrm{Ax}\dfrac{}{p,\Gamma\!\Rightarrow\! p}$\quad$\mathrm{Ax}_\sim\dfrac{}{\simm p,\Gamma\!\Rightarrow\!\simm p}$\quad$\simm\simm_l\dfrac{\phi,\Gamma\!\Rightarrow\!\psi}{\simm\simm\phi,\Gamma\!\Rightarrow\!\psi}$\quad$\simm\simm_r\dfrac{\Gamma\!\Rightarrow\!\phi}{\Gamma\!\Rightarrow\!\simm\simm\phi}$\\[.35em]
$\wedge_l\dfrac{\phi,\chi,\Gamma\!\Rightarrow\!\psi}{\phi\!\wedge\!\chi,\Gamma\!\Rightarrow\!\psi}$\quad$\wedge_r\dfrac{\Gamma\!\Rightarrow\!\phi~~\Gamma\!\Rightarrow\!\chi}{\Gamma\!\Rightarrow\!\phi\!\wedge\!\chi}$\quad$\vee\!_l\dfrac{\phi,\Gamma\!\Rightarrow\!\psi~~\chi,\Gamma\!\Rightarrow\!\psi}{\phi\!\vee\!\chi,\Gamma\!\Rightarrow\!\psi}$\quad$\vee\!_{r_1}\dfrac{\Gamma\!\Rightarrow\!\phi}{\Gamma\!\Rightarrow\!\phi\!\vee\!\chi}$\\[.35em]
$\vee\!_{r_2}\dfrac{\Gamma\!\Rightarrow\!\chi}{\Gamma\!\Rightarrow\!\phi\!\vee\!\chi}$\quad$\simm\!\!\vee\!_l\dfrac{\simm\phi,\simm\chi,\Gamma\!\Rightarrow\!\psi}{\simm(\phi\!\vee\!\chi),\Gamma\!\Rightarrow\!\psi}$\quad$\simm\!\!\vee\!_r\dfrac{\Gamma\!\Rightarrow\!\simm\phi~~\Gamma\!\Rightarrow\!\simm\chi}{\Gamma\!\Rightarrow\!\simm(\phi\!\vee\!\chi)}$\quad$\simm\!\!\wedge\!_{r_1}\dfrac{\Gamma\!\Rightarrow\!\simm\phi}{\Gamma\!\Rightarrow\!\simm(\phi\!\wedge\!\chi)}$\\[.35em]
$\simm\!\!\wedge_{r_2}\dfrac{\Gamma\!\Rightarrow\!\simm\chi}{\Gamma\!\Rightarrow\!\simm(\phi\!\wedge\!\chi)}$\quad$\simm\!\!\vee_l\!\dfrac{\simm\phi,\Gamma\!\Rightarrow\!\psi~~\simm\chi,\Gamma\!\Rightarrow\!\psi}{\sim(\phi\!\wedge\!\chi),\Gamma\!\Rightarrow\!\psi}$\quad$\rightarrow_l\dfrac{\phi\!\rightarrow\!\chi,\Gamma\!\Rightarrow\!\phi~~\chi,\Gamma\!\Rightarrow\!\psi}{\phi\!\rightarrow\!\chi,\Gamma\!\Rightarrow\!\psi}$\\[.35em]
$\rightarrow_r\dfrac{\Gamma,\phi\!\Rightarrow\!\chi}{\Gamma\!\Rightarrow\!\phi\!\rightarrow\!\chi}$\quad$\simm\!\!\rightarrow_l\dfrac{\phi,\simm\chi,\Gamma\!\Rightarrow\!\psi}{\simm(\phi\!\rightarrow\!\chi),\Gamma\!\Rightarrow\!\psi}$\quad$\simm\!\!\rightarrow_r\dfrac{\Gamma\!\Rightarrow\!\phi~~\Gamma\!\Rightarrow\!\simm\chi}{\Gamma\!\Rightarrow\!\simm(\phi\!\rightarrow\!\chi)}$
\end{center}
\caption{Rules of $\Gnfour$}
\label{fig:nfourrules}
\end{figure}

\begin{figure}[!t]
\begin{center}
$\Box\dfrac{\Gamma^\Box\!\Rightarrow\!\chi}{\Gamma\!\Rightarrow\!\Box\chi}$\quad$\lozenge\dfrac{\phi\!\Rightarrow\!\chi}{\Gamma,\lozenge\phi\!\Rightarrow\!\lozenge\chi}$\quad$\Box_\simm\dfrac{\simm\phi\!\Rightarrow\!\simm\chi}{\Gamma,\simm\Box\phi\!\Rightarrow\!\simm\Box\chi}$\quad$\lozenge_\simm\dfrac{\Gamma^\lozenge_\simm\!\Rightarrow\!\simm\chi}{\Gamma\!\Rightarrow\!\simm\lozenge\chi}$\\[.35em]
$\lozenge^\pm\dfrac{\Gamma^\Box,\phi\!\Rightarrow\!\chi}{\Gamma,\lozenge\phi\!\Rightarrow\!\lozenge\chi}$\quad$\Box^\pm_\simm\dfrac{\Gamma^\lozenge_\simm,\simm\phi\!\Rightarrow\!\simm\chi}{\Gamma,\simm\Box\phi\!\Rightarrow\!\simm\Box\chi}$\quad$\lozenge^\curlyvee\dfrac{\Gamma^\lozenge_\simm,\phi\!\Rightarrow\!\chi}{\Gamma,\lozenge\phi\!\Rightarrow\!\lozenge\chi}$\quad$\Box^\curlyvee_\simm\dfrac{\Gamma^\Box,\simm\phi\!\Rightarrow\!\simm\chi}{\Gamma,\simm\Box\phi\!\Rightarrow\!\simm\Box\chi}$\\[.35em]
$\Box^{\Join}\dfrac{\Gamma^\Box,\Gamma^\lozenge_\simm\!\Rightarrow\!\chi}{\Gamma\!\Rightarrow\!\Box\chi}$\quad$\lozenge^{\Join}\dfrac{\simm\phi\!\Rightarrow\!\chi}{\Gamma,\simm\Box\phi\!\Rightarrow\!\lozenge\chi}$\quad$\Box^{\Join}_\simm\dfrac{\phi\!\Rightarrow\!\simm\chi}{\Gamma,\lozenge\phi\!\Rightarrow\!\simm\Box\chi}$\quad$\lozenge^{\Join}_\simm\dfrac{\Gamma^\Box,\Gamma^\lozenge_\simm\!\Rightarrow\!\simm\chi}{\Gamma\!\Rightarrow\!\simm\lozenge\chi}$\\[.35em]
$\lozenge^1\dfrac{\Gamma^\Box,\Gamma^\lozenge_\simm,\phi\!\Rightarrow\!\chi}{\Gamma,\lozenge\phi\!\Rightarrow\!\lozenge\chi}$\quad$\Box^1_\simm\dfrac{\Gamma^\Box,\Gamma^\lozenge_\simm,\simm\phi\!\Rightarrow\!\simm\chi}{\Gamma,\simm\Box\phi\!\Rightarrow\!\simm\Box\chi}$\quad$\lozenge^{1,\Join}\dfrac{\Gamma^\Box,\Gamma^\lozenge_\simm,\simm\phi\!\Rightarrow\!\chi}{\Gamma,\simm\Box\phi\!\Rightarrow\!\lozenge\chi}$\quad$\Box^{1,\Join}_\simm\dfrac{\Gamma^\Box,\Gamma^\lozenge_\simm,\phi\!\Rightarrow\!\simm\chi}{\Gamma,\lozenge\phi\!\Rightarrow\!\simm\Box\chi}$
\end{center}
\caption{Modal rules for $\nfourck$-style logics}
\label{fig:modal-rules}
\end{figure}
\begin{definition}\label{def:GCN4Kextensions}
A sequent is an expression $\Gamma\Rightarrow\phi$, where $\Gamma$ is a~finite multi-set of formulas and $\phi$ is a formula. The \emph{formula interpretation} of $\Gamma\Rightarrow\phi$ is the formula $\bigwedge_{\psi\in\Gamma}\psi\rightarrow\phi$. A~sequent is $\Lmsf$-\emph{valid} if so is its formula interpretation.

Given a~multiset of formulas $\Gamma$, we let $\Gamma^\Box=\{\phi\mid\square\phi\in\Gamma\}$ and $\Gamma^\lozenge_\sim=\{\simm\phi\mid\simm\lozenge\phi\in\Gamma\}$. 
We define the following calculi:
\begin{align*}
\cnfck&=\Gnfour\oplus\{\Box,\lozenge,\Box_\sim,\lozenge_\sim\}\quad\cnfckbd=\Gnfour\oplus\{\Box,\lozenge^\curlyvee,\Box^\curlyvee_\sim,\lozenge_\sim\}\\
\cnfckpm&=\Gnfour\!\oplus\!\{\Box,\lozenge^\pm,\Box^\pm_\sim,\lozenge_\sim\}\quad\cnfckj\!=\!\Gnfour\!\oplus\!\{\Box^{\Join},\lozenge,\lozenge^{\Join},\Box_\sim,\Box^{\Join}_\sim,\lozenge^{\Join}_\sim\}\\
\cnfckone&=\Gnfour\!\oplus\!\{\Box^{\Join},\lozenge^1,\lozenge^{1,\Join},\Box^{1}_\sim,\Box^{1,\Join}_\sim,\lozenge^{\Join}_\sim\}
\end{align*}
Propositional rules which constitute $\Gmc\nfour$ are found in Figure \ref{fig:nfourrules} and modal rules are found in Figure \ref{fig:modal-rules}.
\end{definition}

As usual, we define a~\emph{$\Gmc\Lmsf$-derivation} as an upwards-branching tree of sequents s.t.\ each node is obtained from its parents via an application of a~rule from $\Gmc\Lmsf$. A~\emph{proof} is a~finite derivation whose leaves are axioms. The \emph{height} of a sequent in a finite derivation is the height of the corresponding node in the derivation (that is the height of the subtree rooted in that node).
\begin{definition}
A rule is \emph{height-preserving (hp) invertible} if, in case the conclusion has a proof of height at most $n$, so do all its premises. 
A rule is \emph{height-preserving (hp) admissible} if, in case the premises have a proof of height at most $n$, so does the conclusion. 
\end{definition}

We can show by induction on the complexity of a formula that 
\begin{proposition}\label{prop:axioms}
For any $\phi$ and  $\Gamma$, $\seq{\Gamma,\phi}{\phi}$ is provable in $\Gmc\Lmsf$. 
\end{proposition}

Note that the context changes from the conclusion of a~modal rule to its premise depending on which accessibility relations coincide. E.g., as in $\nfourck$ all four relations are distinct, the $\Box$ rule only proceeds from $\Gamma$ to $\Gamma^\Box$. On the other hand, as $R^+_\Box=R^-_\lozenge$ in $\Join$-birelational frames, $\Box^{\Join}$ proceeds from $\Gamma$ to $\Gamma^\Box\cup\Gamma^\lozenge_\sim$. 

We show that axioms $\Join_\Box$ and $\curlyvee_\lozenge$ are provable in $\cnfckj$ and $\cnfckbd$, 
\begin{center}
\begin{adjustbox}{max width=\textwidth}
\begin{tabular}{ccc}
\AXC{}
\LeftLabel{\tiny Prop \ref{prop:axioms}}
 \UIC{$\seq{\phi}{\phi}$}
 \LeftLabel{$\simm\simm_r$}
 \UIC{$\seq{\phi}{\simm\simm\phi}$}
 \LeftLabel{$\Diamondjoin_\simm$}
 \UIC{$\seq{\Box\phi}{\simm\lozenge\simm\phi}$}
 \LeftLabel{$\to_r$}
 \UIC{$\seq{}{\Box\phi\to\simm\lozenge\simm\phi}$}
 \AXC{}
 \LeftLabel{\tiny Prop \ref{prop:axioms}}
 \UIC{$\seq{\phi}{\phi}$}
 \LeftLabel{$\simm\simm_l$}
 \UIC{$\seq{\simm\simm\phi}{\phi}$}
 \LeftLabel{$\Boxjoin$}
 \UIC{$\seq{\simm\lozenge\simm\phi}{\Box\phi}$}
 \LeftLabel{$\to_r$}
 \UIC{$\seq{}{\simm\lozenge\simm\phi\to\Box\phi}$}
 \LeftLabel{$\wedge_r$}
 \BIC{$\seq{}{(\Box\phi\to\simm\lozenge\simm\phi)\wedge (\simm\lozenge\simm\phi\to\Box\phi)}$}
 \DP
 &
 \qquad\qquad
 &
 \AXC{}
 \LeftLabel{\tiny Prop \ref{prop:axioms}}
 \UIC{$\seq{\phi\to\chi,\phi}{\phi}$}
 \AXC{}
 \LeftLabel{\tiny Prop \ref{prop:axioms}}
 \UIC{$\seq{\phi,\chi}{\chi}$}
 \LeftLabel{$\to_l$}
 \BIC{$\seq{\phi\to\chi,\phi}{\chi}$}
 \LeftLabel{$\simm\simm_l$}
 \UIC{$\seq{\simm\simm(\phi\to\chi),\phi}{\chi}$}
 \LeftLabel{$\Diamondbd$}
 \UIC{$\seq{\simm\lozenge\simm(\phi\to\chi),\lozenge\phi}{\lozenge\chi}$}
 \LeftLabel{$\to_r\times 2$}
 \UIC{$\seq{}{\simm\lozenge\simm(\phi\to\chi)\to (\lozenge\phi\to\lozenge\chi)}$}
 \DP
 \end{tabular}
 \end{adjustbox}
 \end{center}

The soundness of calculi can be shown by checking that the validity of sequents is preserved from the premises to the conclusion of every rule.
\begin{restatable}{theorem}{rule-validity}\label{theorem:rule-validity}
For any $\Lmsf$, it holds that $\Gmc\Lmsf\vdash\Gamma\Rightarrow\phi$ implies $\Lmsf\models\bigwedge_{\psi\in\Gamma}\psi\rightarrow\phi$.
\end{restatable}

We prove the completeness of $\Gmc\Lmsf$ calculi in the following section as a~consequence of the admissibility of the $\cut$ rule.
\begin{center}
 \AXC{$\seq{\Gamma}{\phi}$}
 \AXC{$\seq{\phi,\Delta}{\chi}$}
 \LeftLabel{$\cut$}
 \BIC{$\seq{\Gamma,\Delta}{\chi}$}
 \DP
\end{center}

%% file: 05-cut.tex
\section{Cut-admissibility}

In order  to prove  the admissibility of cut for all systems, we first observe that the calculus $\Gmc\nfour$ in Figure~\ref{fig:nfourrules} is a variant of the one for logic $\nfour$ provided in~\cite{Lopez-Escobar1972,KamideWansing2012}. For convenience, we refer to the latter as $\Cmc\nfour$. To compare the two calculi, in $\Cmc\nfour$, structural rules like weakening and contraction are primitive rules, initial sequents (or `axioms') are simply $\seq{p}{p}$ and $\seq{\simm p}{\simm p}$ and $\to_l$ is of the following form where the implication formula is no longer kept in the left premise and the two premises have independent contexts $\Gamma$ and $\Delta$:
\begin{prooftree}
 \AXC{$\seq{\Gamma}{\phi}$}
 \AXC{$\seq{\chi,\Delta}{\psi}$}
 \LeftLabel{$\to_l$}
 \BIC{$\seq{\phi\to\chi,\Gamma,\Delta}{\psi}$}
\end{prooftree}
As a result, $\Cmc\nfour$ is shown to admit cut-elimination but is neither weakening-free nor contraction-free. However, with the modification on $\to_l$, we can show that both rules are \emph{hp admissible} in $\Gmc\nfour$. This relies on the fact that the modified $\to_l$ is height-preserving invertible on the right premise and all the other left rules in $\Gmc\nfour$ are hp invertible. 
Besides, all the right rules except $\vee_{r_i}$ and $\simm\wedge_{r_i}$ are also 
hp invertible. 
We will use this property to prove cut-admissibility and obtain decidability for the logics we consider. The proof for the basic $\Gmc\nfour$ relies on cut-elimination of $\Cmc\nfour$ (see the appendix).
\begin{proposition}\label{prop:weakening-contraction-admissibility}
    The following rules are height-preserving admissible in $\Gmc\Lmsf$:
    \begin{center}
        \AXC{$\seq{\Gamma}{\chi}$}
        \LeftLabel{\textsf{weakening}}
        \UIC{$\seq{\phi,\Gamma}{\chi}$}
        \DP
        \hfil
        \AXC{$\seq{\phi,\phi,\Gamma}{\chi}$}
        \LeftLabel{\textsf{contraction}}
        \UIC{$\seq{\phi,\Gamma}{\chi}$}
        \DP
    \end{center}
\end{proposition}

Let us compare rules in Figure \ref{fig:modal-rules}. We say that $\Rmc$ is \textit{weaker} than $\Rmc'$ ($\Rmc\lessdot\Rmc'$) if $\Rmc$ and $\Rmc'$ have the same conclusion while the premise of $\Rmc$ is a subset (in the sense of multiset) of that of $\Rmc'$. We have the following relations: 
\begin{align*}
\Box\lessdot\Boxjoin&&\Boxjoin_\simm\lessdot\Boxonejoin_\simm&&\lozenge\lessdot\Diamondbd\lessdot\lozenge^1&&\Box_\simm\lessdot\Boxpmsim\lessdot\Boxonesim\\
\lozenge_\simm\lessdot\Diamondjoin_\simm&&\Diamondjoin\lessdot\Diamondonejoin&&\lozenge\lessdot\Diamondpm\lessdot\lozenge^1&&\Box_\simm\lessdot\Boxbdsim\lessdot\Boxonesim
\end{align*}
The `weaker' rules are derivable in calculi containing the `stronger' rules by applying weakening to the premises. 






\begin{definition}\label{def:occurrences}
In a $\Gmc\Lmsf$-derivation, we say an occurrence of a~formula in the conclusion of an application of $\Rmc$ is
\emph{side} if it is irrelevant to the rule application and also kept in one of the premises;
\emph{weak} if it is irrelevant to the rule application and does not occur in any of the premises;
\emph{principal}, otherwise.
\end{definition}

For example, in $\wedge_r$, formulas in $\Gamma$ occur as side formulas while $\phi\wedge \chi$ is principal; in $\to_l$, formulas in $\Gamma$ and formula $\psi$ occur as side formulas while $\phi\rightarrow\chi$ is principal. In modal rules like $\Diamondonejoin$, formula $\lozenge\chi,\simm\Box\phi$ and all the $\Box$ or $\simm\lozenge$-prefixed formulas in $\Gamma$ are principal while other formulas in $\Gamma$ are weak. No side formulas occurs in any application of modal rules.

\begin{restatable}[Cut-admissibility]{theorem}{cutadmissibility}\label{theorem:cutadmissibility}
The $\cut$ rule is admissible in $\Gmc\Lmsf$. 
\end{restatable}

\begin{proof}
We provide the proof for $\cnfckone$. For other systems, a proof can be obtained by taking a subset of cases presented in this proof, as all the other rules from Figure \ref{fig:modal-rules} are derivable in $\cnfckone$ with applications of weakening. 


For each application of $\cut$, we associate a pair $(h,d)$ as its \textit{rank}, where $h$ denotes the sum of the heights of the two premise of $\cut$ in the derivation and $d$ denotes complexity of the cut formula. We proceed by double induction on the rank of $\cut$. Consider a derivation $\Dmc$ in $\cnfckone$ that has the form 
 \begin{center}
 \vspace{-.3cm}
 $
 \vlderivation{
 \vliin{}{\cut}{\seq{\Gamma,\Delta}{\chi}}{
 \vltr {\Dmc_1} {
 \seq{\Gamma}{\phi}
 }{
 \vlhy {\quad}}
 {
 \vlhy {}}
 {
 \vlhy {\quad}}
 }
 {
 \vltr {\Dmc_2} {
 \seq{\phi,\Delta}{\chi}
 }{
 \vlhy {\quad}}
 {
 \vlhy {}}
 {
 \vlhy {\quad}} 
 }
 }
 $
 \end{center}
We denote the last rule application in $\Dmc_1$ and $\Dmc_2$ as $\Rmc_1$ and $\Rmc_2$ respectively and discuss possible different roles of the cut formula $\phi$ in $\Rmc_1$ and $\Rmc_2$. We omit details in the cases when both $\Rmc_1$ and $\Rmc_2$ are $\nfour$ rules as they are covered by the cut-admissibility of $\Gmc\nfour$, for which the proof can be found in the appendix. 

Given a cut formula $\phi$, we have four cases. 
(1) one of the premises is an axiom; 
(2)
$\phi$ is principal in \textit{both} premises; 
(3) 
$\phi$ is principal in the left premise \textit{only};
(4) 
$\phi$ is \emph{not principal} in the left premise. We only consider the second case here while other cases are shown in the appendix. 

If $\phi$ is a propositional formula, all cases are part of the cut-elimination of $\Gmc\nfour$. If $\phi=\Box\psi$, we have $\Rmc_1=\Boxjoin$ and $\Rmc_2$ is any modal rule. All cases can be done by applying $\cut$ first and then a suitable modal rule. We only provide one example with $\Rmc_1=\Boxjoin$ and $\Rmc_2=\Boxonesim$ as other cases are similar. In this case, $\chi$ is some $\simm\Box\theta$ and $\Delta=\Sigma,\simm\Box\mu$. We unfold $\Dmc$ and transform it as follows.
\begin{center}
\begin{adjustbox}{max width=.9\textwidth}
\begin{tabular}{ccc}
$\vlderivation{\vliin{\cut}{}{\seq{\Gamma,\Sigma,\simm\Box\mu}{\simm\Box\theta}}{\vlin{\Boxjoin}{}{\seq{\Gamma}{\Box\psi}}{\vltr{\Dmc_1'}{\seq{\Gammabox,\Gammadianeg}{\psi}}{\vlhy{\quad}}{\vlhy{}}{\vlhy{\quad}}}}{\vlin{\Boxonesim}{}{\seq{\Box\psi,\Sigma,\simm\Box\mu}{\simm\Box\theta}}{\vltr{\Dmc_2'}{\seq{\psi,\Sigma^\Box,\diasimm{\Sigma},\simm\mu}{\simm\theta}}{\vlhy{\quad}}{\vlhy{}}{\vlhy{\quad}}}}}$
&\quad$\dashrightarrow$\quad&
$\vlderivation{\vlin{\Boxonesim}{}{\seq{\Gamma,\Sigma,\simm\Box\mu}{\simm\Box\theta}}{\vliin{\cut}{}{\seq{\Gamma^\Box,\Gammadianeg,\Sigma^\Box,\diasimm{\Sigma},\simm\mu}{\simm\theta}}{\vltr{\Dmc_1'}{\seq{\Gamma^\Box,\Gammadianeg}{\psi}}{\vlhy{\quad}}{\vlhy{}}{\vlhy{\quad}}}{\vltr{\Dmc_2'}{\seq{\psi,\Sigma^\Box,\diasimm{\Sigma}, \simm\mu}{\simm\theta}}{\vlhy {\quad}}{\vlhy{}}{\vlhy{\quad}}}}}$
\end{tabular}
\end{adjustbox}
\end{center}
If $\phi=\lozenge\psi$, then $\Rmc_1\in\{\lozenge^1,\Diamondonejoin\}$ and $\Rmc_2\in \{\lozenge^1,\Boxonejoin\}$. The proof for all four combinations proceeds as in the case of $\phi=\Box\psi$. The cases when $\phi=\simm\Box\psi$ and $\phi=\simm\lozenge\psi$ are dual to the cases of $\lozenge\psi$ and $\Box\psi$, respectively.
\qed
\end{proof}

Completeness of sequent calculi, decidability of $\nfourck$ and its extensions, and disjunctive and constructive falsity properties now follow by standard arguments from Theorem~\ref{theorem:cutadmissibility}.
\begin{restatable}{theorem}{sequentcompleteness}\label{theorem:sequentcompleteness}
$\Lmsf\models\bigwedge_{\psi\in\Gamma}\psi\rightarrow\phi$ implies $\Gmc\Lmsf\vdash\Gamma\Rightarrow\phi$.
\end{restatable}
\begin{theorem}\label{theorem:nfourckdecidability}
Validity in $\nfourck$ and its extensions is decidable.
\end{theorem}
\begin{proof}
Since contraction is admissible in all calculi we can consider their set-based formulation where in a sequent $\Gamma\Rightarrow\phi$, $\Gamma$ is a set rather than a multiset. All rules are ‘weakly’ analytical in the sense that for every rule $\Rmc$ with conclusion $\Gamma\Rightarrow\phi$, all formulas occurring in the premises of $\Rmc$ are either subformulas of formulas in $\Gamma\cup\{\phi\}$ or obtained by prefixing a negation-free subformula in this set with at most two $\simm$. Therefore if we consider any derivation $\Dmc$ with a root sequent $\Gamma\Rightarrow\chi$, there are only finitely many distinct sequents that can occur in $\Dmc$ (as they can contain only a subset of subformulas of $\Gamma\cup\{\chi\}$ and their negations $\simm$). We can w.l.o.g.\ restrict the proof search to derivations (let us call them ‘regular’) such that no branch contains the same sequent twice. As a~consequence, given a~sequent $\Gamma\Rightarrow\chi$, there are only finitely many regular derivations with root $\Gamma\Rightarrow\chi$. Thus the (obvious) decision procedure for $\Gamma\Rightarrow\chi$ generates subsequently each regular derivation $\Dmc$ testing whether it is a proof or not and reporting Success as soon as it finds a~proof or Failure after having exhausted all derivations without finding a proof. 
\qed
\end{proof}
\begin{theorem}\label{theorem:constructivness}
Let $\Lmsf\in\{\nfourck,\nfourck^\curlyvee,\nfourck^\pm,\nfourck^{\Join},\nfourck^1\}$. Then:
\begin{align*}
\Lmsf\models\phi\vee\chi&\text{ iff }\Lmsf\models\phi\text{ or }\Lmsf\models\chi&\Lmsf\models\simm(\phi\wedge\chi)&\text{ iff }\Lmsf\models\simm\phi\text{ or }\Lmsf\models\simm\chi
\end{align*}
\end{theorem}

%% file: 06-conclusion.tex
\section{Conclusion and perspectives\label{sec:conclusion}}

There are several axes of further research.\hide{ From a computational viewpoint, } First, even though we established the decidability of $\nfourck$ logics, \hide{by means of the proposed Gentzen cut-free calculi, }their exact complexity remains open. We conjecture that all our logics are $\pspace$-complete. It would be also instructive to construct analytic calculi that \hide{also }produce \emph{counter-models} witnessing non-validity.

From a semantical perspective, a natural development 
would be the study of other Nelson-like paraconsistent modal logics such as those presented in~\cite{Sherkhonov2008} and~\cite{OdintsovWansing2003,OdintsovWansing2008}. 
In particular, we are interested in establishing a~hierarchy of paraconsistent constructive logics corresponding to the frame conditions connecting accessibility relations and preordering of the states.

Another \hide{obvious} extension would be to consider the usual properties of the accessibility relations that characterise the epistemic, doxastic, or deontic interpretation of the modalities. In all cases, the study of semantic and axiomatic extensions should be accompanied by the development of analytic proof systems which can provide a decision procedure for the respective logics.

%% file: appendix.tex
\appendix
\section*{Appendix}
\trivialmodel*

\truthlemmabasic*
\begin{proof}
We consider the remaining cases of modal formulas: $\phi=\Box\chi$, $\phi=\simm\Box\chi$, and $\phi=\lozenge\chi$. First, let $\phi=\Box\chi$. Again, we assume for contradiction that $\Mfrak^\Cmsf,\sfrak\Vdash^+\Box\chi$ but $\Box\chi\notin\hmbf(\sfrak)$. Consider $\Xi^!=\{\tau\mid\Box\tau\in\hmbf(\sfrak)\}$. It is clear that $\Xi^!\nvdash_{\Hnfourck}\chi$ because otherwise we would have $\Hnfourck\vdash\bigwedge\Xi'\rightarrow\chi$ for some finite subset of $\Xi^!$, whence, $\Hnfourck\vdash\Box\bigwedge\limits_{\tau'\in\Xi'}\tau'\rightarrow\Box\chi$ by $\rmbf_\Box$ and $\Hnfourck\vdash\bigwedge\limits_{\tau'\in\Xi'}\Box\tau'\rightarrow\Box\chi$ using Lemma~\ref{lemma:modalitydistribution}. But then, we would have that $\{\Box\tau'\mid\tau'\in\Xi'\}\vdash_{\Hnfourck}\Box\chi$ which contradicts the assumption that $\Box\chi\notin\hmbf(\sfrak)$. Thus, again, using Proposition~\ref{prop:saturatednoderivation}, we extend $\Xi^!$ to a~saturated set~$\Xi^\sharp$ s.t.\ $\Xi^\sharp\nvdash_{\Hnfourck}\chi$ and let $\sfrak'=\langle\Xi,\{\Xi^\sharp\},\Phi^-_\Box,\Phi^+_\lozenge,\Phi^-_\lozenge\rangle$. Now let $\sfrak''$ be a~segment s.t.\ $\hmbf(\sfrak'')=\Xi^\sharp$. Clearly, $\sfrak'{R^+_\Box}^\Cmsf\sfrak''$ and by the induction hypothesis, $\Mfrak^\Cmsf,\sfrak''\nVdash^+\chi$. But as $\sfrak\leq^\Cmsf\sfrak'$, we have that $\Mfrak^\Cmsf,\sfrak\nVdash^+\Box\chi$, contrary to the assumption.

Conversely, let $\Box\chi\in\hmbf(\sfrak)$. It follows that $\Box\chi\in\hmbf(\sfrak')$ for every $\sfrak'\geq^\Cmsf\sfrak$. But then $\chi\in\hmbf(\sfrak'')$ for every $\sfrak''\in{R^+_\Box}^\Cmsf(\sfrak')$. Hence, by the induction hypothesis, we have that $\Mfrak^\Cmsf,\sfrak''\Vdash^+\chi$ for every such $\sfrak''$, as required.

Let $\phi=\simm\Box\chi$. We show that $\simm\Box\chi\in\hmbf(\sfrak)$ iff $\Mfrak^\Cmsf,\sfrak\Vdash^-\Box\chi$. Assume for contradiction that $\Mfrak^\Cmsf,\sfrak\Vdash^-\Box\chi$ but $\simm\Box\chi\notin\hmbf(\sfrak)$. Similarly to the case of $\lozenge$, we can see that $\simm\varrho\nvdash_{\Hnfourck}\simm\chi$ for every $\simm\Box\varrho\notin\hmbf(\sfrak)$. Indeed, otherwise, we would infer $\simm\Box\varrho\vdash_{\Hnfourck}\simm\Box\chi$ (whence, $\hmbf(\sfrak)\vdash_{\Hnfourck}\simm\Box\chi$, contrary to the assumption) using $\rmbf^\sim_\Box$ and the deduction theorem. Now, we can extend $\{\simm\varrho\}$ to a~saturated set $\Xi_{\simm\varrho}$ s.t.\ $\Xi_{\simm\varrho}\nvdash_{\Hnfourck}\simm\chi$ and see that $\sfrak'=\left\langle\Xi,\Phi^+_\Box,\{\Xi_{\simm\varrho}\mid\simm\Box\varrho\in\Xi\},\Phi^+_\lozenge,\Phi^-_\lozenge\right\rangle$
is a~segment s.t.\ $\sfrak'\geq^\Cmsf\sfrak$. Moreover, $\sfrak'{R^-_\Box}^\Cmsf\sfrak''$ implies that $\hmbf(\sfrak'')=\Xi_{\simm\varrho}$ for some $\simm\varrho$. As $\simm\chi\notin\Xi_{\simm\varrho}$ for every $\Xi_{\simm\varrho}$, we have that $\Mfrak^\Cmsf,\sfrak''\nVdash^+\simm\chi$ (whence, $\Mfrak^\Cmsf,\sfrak''\nVdash^-\chi$) for every $\sfrak''\in{R^-_\Box}^\Cmsf(\sfrak')$. This, however, contradicts the assumption that $\Mfrak^\Cmsf,\sfrak\Vdash^-\Box\chi$.

Conversely, let $\simm\Box\chi\in\hmbf(\sfrak)$. Thus, for every $\sfrak'\geq^\Cmsf\sfrak$, we have $\simm\Box\chi\in\hmbf(\sfrak')$, whence, $\simm\chi\in{\Phi^-_\Box}'$. This means that there is some $\sfrak''$ s.t.\ $\hmbf(\sfrak'')={\Phi^-_\Box}'$ (i.e., $\sfrak'{R^-_\Box}^\Cmsf\sfrak''$) and $\simm\chi\in\hmbf(\sfrak'')$. By the induction hypothesis, we obtain that $\Mfrak^\Cmsf,\sfrak''\Vdash^+\simm\chi$ and hence, $\Mfrak^\Cmsf,\sfrak''\Vdash^-\chi$, as required.

Let $\phi=\lozenge\chi$. Assume for contradiction that $\Mfrak^\Cmsf,\sfrak\Vdash^+\lozenge\chi$ but $\lozenge\chi\notin\hmbf(\sfrak)$. Now, for $\lozenge\varrho\in\Xi$, one can see that $\varrho\nvdash_{\Hnfourck}\chi$, for otherwise we would infer $\lozenge\varrho\vdash_{\Hnfourck}\lozenge\chi$ by $\rmbf_\lozenge$ and the deduction theorem, whence, $\Xi\vdash_{\Hnfourck}\lozenge\chi$, contrary to the assumption. Hence, using Proposition~\ref{prop:saturatednoderivation} we can extend $\{\varrho\}$ to a~saturated set $\Xi_\varrho$ s.t.\ $\Xi_\varrho\nvdash_{\Hnfourck}\chi$. Clearly, $\sfrak'\!=\!\left\langle\Xi,\Phi^+_\Box,\Phi^-_\Box,\{\Xi_\varrho\mid\lozenge\varrho\in\Xi\},\Phi^-_\lozenge\right\rangle$
is a~segment and $\sfrak\leq^\Cmsf\sfrak'$. Moreover, $\sfrak'{R^+_\lozenge}^\Cmsf\sfrak''$ implies that $\hmbf(\sfrak'')=\Xi_\varrho$ for some $\varrho$. As $\chi\notin\Xi_\varrho$ (for every $\Xi_\varrho$), we have by the application of the induction hypothesis, that $\Mfrak^\Cmsf,\sfrak''\nVdash^+\chi$ for every $\sfrak''\in{R^+_\lozenge}^\Cmsf(\sfrak')$. But this contradicts the assumption that $\Mfrak^\Cmsf,\sfrak\Vdash^+\lozenge\chi$ as now $\sfrak'$ is the required $\leq^\Cmsf$-successor of $\sfrak$ s.t.\ $\chi$ is \emph{not true} in any of its ${R^+_\lozenge}^\Cmsf$-successors.

Conversely, let $\lozenge\chi\in\hmbf(\sfrak)$. Then for every $\sfrak'\geq^\Cmsf\sfrak$, we have that $\lozenge\chi\in\hmbf(\sfrak')$, whence $\chi\in{\Phi^+_\lozenge}'$, and thus, there is some $\sfrak''$ s.t.\ $\hmbf(\sfrak'')={\Phi^+_\lozenge}'$ (hence, $\sfrak'{R^+_\lozenge}^\Cmsf\sfrak''$) and $\chi\in\hmbf(\sfrak'')$. By the induction hypothesis, we get $\Mfrak^\Cmsf,\sfrak''\Vdash^+\chi$, as required.\qed
\end{proof}
\truthlemmaextensions*
\begin{proof}
The only detail in the proof of Lemma~\ref{lemma:truthlemmabasic} that we need to change is the treatment of modalities. For $\Hnfourck^{\Join}$, the difference is in the treatment of the $\Box\chi$ and $\simm\lozenge\chi$ cases. Namely, we set $\Xi^!=\{\tau\mid\Box\tau\in\Xi\}\cup\{\simm\sigma\mid\simm\lozenge\sigma\in\Xi\}$. Again, it is clear that $\Xi\nvdash_{\Hnfourck^{\Join}}\chi$. Otherwise, we would have $\Xi'\vdash_{\Hnfourck^{\Join}}\chi$ for some finite $\Xi'\subseteq\Xi^!$. But then $\Hnfourck^{\Join}\vdash\left(\bigwedge\limits_{\tau'\in\Xi'}\!\!\!\!\Box\tau'\wedge\bigwedge\limits_{\simm\sigma'\in\Xi'}\!\!\!\!\Box\simm\sigma'\right)\rightarrow\Box\chi$
using $\rmbf_\Box$ and Lemma~\ref{lemma:modalitydistribution}, whence, applying $\Join_\Box$ and $\rmbf^\sim_\lozenge$ which give us $\Box\simm\sigma'\leftrightarrow\simm\lozenge\simm\simm\sigma'$ and $\simm\lozenge\simm\simm\sigma'\!\!\leftrightarrow\!\!\simm\lozenge\sigma'$, we would get $\Hnfourck^{\Join}\!\vdash\!\left(\bigwedge\limits_{\tau'\in\Xi'}\!\!\!\!\Box\tau'\wedge\bigwedge\limits_{\simm\sigma'\in\Xi'}\!\!\!\!\!\!\simm\lozenge\sigma'\right)\!\rightarrow\!\Box\chi$
contrary to the assumption. Note, alternatively, that due to the deductive closure of $\Xi$, the two halves of $\Xi^!$ coincide. This can be seen by showing that $\Hnfourck^{\Join}\vdash\simm\lozenge\psi\leftrightarrow\Box\simm\psi$ using $\rmbf^\sim$ and $\Join_\Box$. Thus, if $\Xi^\sharp$ is a~saturation of $\Xi^!$ s.t.\ $\Xi^\sharp\nvdash_{\Hnfourck^{\Join}}\chi$, $\sfrak'=\langle\Xi,\Xi^\sharp,\Phi^-_\Box,\Phi^+_\lozenge,\Xi^\sharp\rangle$ and $\hmbf(\sfrak'')=\Xi^\sharp$, it follows that $\sfrak'{R^+_\Box}^\Cmsf\sfrak''$ and $\sfrak'{R^-_\lozenge}^\Cmsf\sfrak''$. Whence, $\Mfrak^\Cmsf,\sfrak\nVdash^+\Box\chi$, as required.

The proof of the $\simm\lozenge\chi$ case is similar. The rest of the proof is the same as for $\Hnfourck$. Note also that it is possible to treat $\lozenge$ as a~shorthand for $\simm\Box\simm$ and provide a~mono-modal axiomatisation of $\nfourck^1$.

Let us now consider the case of $\Hnfourck^\pm$. Assume that $\sfrak$ is a~$\pm$-segment. We show that $\lozenge\chi\in\hmbf(\sfrak)$ iff $\Mfrak^\Cmsf,\sfrak\Vdash^+\lozenge\chi$. Suppose for contradiction that $\Mfrak^\Cmsf,\sfrak\Vdash^+\lozenge\chi$ but $\lozenge\chi\notin\hmbf(\sfrak)$. For $\lozenge\varrho\in\Xi$, consider $\Xi^\pm_\varrho=\{\varrho\}\cup\{\tau\mid\Box\tau\in\Xi\}$ and observe that $\Xi^\pm_\varrho\nvdash_{\Hnfourck^\pm}\chi$ (the reasoning is the same as in Lemma~\ref{lemma:truthlemmabasic}). Thus, we extend $\Xi^\pm_\varrho$ to a~saturated $\Xi'_\varrho$ s.t.\ $\Xi'_\varrho\nvdash_{\Hnfourck^\pm}\chi$ and consider $\sfrak'=\langle\Xi,\{\Xi'_\varrho\mid\lozenge\varrho\in\Xi\},\Phi^-_\Box,\{\Xi'_\varrho\mid\lozenge\varrho\in\Xi\},\Phi^-_\lozenge\rangle$.
It is easy to see that $\sfrak'$ is a~segment s.t.\ $\sfrak'\geq^\Cmsf\sfrak$. The proof of the rest of the case is the same as for $\Hnfourck$. The case of $\simm\Box\chi$ is dual: we consider the saturation of $\Xi^\pm_{\simm\varrho}=\{\simm\varrho\}\cup\{\simm\tau\mid\simm\lozenge\tau\in\Xi\}$.

The main difference in the case of $\Hnfourck^\curlyvee$ is the treatment of $\lozenge$ and $\simm\Box$. For every $\lozenge\varrho\in\Xi$, we consider the saturation of $\Xi^\curlyvee_\varrho=\{\varrho\}\cup\{\simm\tau\mid\simm\lozenge\tau\in\Xi\}$. Dually, for every $\simm\Box\varrho\in\Xi$, we consider the saturation of $\Xi^\curlyvee_{\simm\varrho}=\{\simm\varrho\}\cup\{\tau\mid\Box\tau\in\Xi\}$. The rest of the proof is the same as for $\Hnfourck$.
\qed
\end{proof}
\cutadmissibility*
\begin{proof}
We first show the cut-admissibility for $\Gmc\nfour$. We establish the proof by transforming an arbitrary derivation in $\Gmc\nfour+\cut$ into another one in $\Gmc\nfour$ without applications of $\cut$. The transformation is further divided into the following steps:
(1) from $\Gmc\nfour+\cut$ to $\Cmc\nfour+\cut$; 
(2) from $\Cmc\nfour+\cut$ to $\Cmc\nfour$;
(3) from $\Cmc\nfour$ to $\Gmc\nfour$. 
 (2) is due to the fact that $\Cmc\nfour$ has cut-admissibility \cite{KamideWansing2012}, thus it suffices to show (1) and (3). 
 For (1), to transform a derivation $\Dmc$ in $\Gmc\nfour+\cut$, we keep all the other rule applications and only need to modify axioms and $\to_l$. Recall weakening and contraction are primitive rules in $\Cmc\nfour$, thus we have 
 \begin{center}
 \begin{adjustbox}{max width=.9\textwidth}
 \begin{tabular}{ccc}
 $\seq{(\simm) p,\Gamma}{(\simm) p}$
 &
 \quad $\dashrightarrow$ \quad
 &
 \AXC{$\seq{(\simm)p}{(\simm)p}$}
 \LeftLabel{$w$}
 \UIC{$\seq{(\simm) p,\Gamma}{(\simm) p}$}
 \DP
 \\[.5cm]
 \AXC{$\seq{\phi\to\chi,\Gamma}{\phi}$}
 \AXC{$\seq{\chi,\Gamma}{\psi}$}
 \LeftLabel{$\to_l$}
 \BIC{$\seq{\phi\to\chi,\Gamma}{\psi}$}
 \DP
 &
 \quad $\dashrightarrow$ \quad
 &
 \AXC{$\seq{\phi\to\chi,\Gamma}{\phi}$}
 \AXC{$\seq{\chi,\Gamma}{\psi}$}
 \LeftLabel{$\to_l$}
 \BIC{$\seq{\phi\to\chi,\phi\to\chi,\Gamma}{\psi}$}
 \LeftLabel{$c$}
 \UIC{$\seq{\phi\to\chi,\Gamma}{\psi}$}
 \DP
 \end{tabular}
 \end{adjustbox}
 \end{center}

For (3), note that initial sequents in $\Cmc\nfour$ are also initial in $\Gmc\nfour$, and for an application of $\to_l$, we have 
\begin{center}
\begin{adjustbox}{max width=.9\textwidth}
\begin{tabular}{ccc}
 \AXC{$\seq{\Gamma}{\phi}$}
 \AXC{$\seq{\chi,\Delta}{\psi}$}
 \LeftLabel{$\to_l$}
 \BIC{$\seq{\phi\to\chi,\Gamma,\Delta}{\psi}$}
 \DP
 &
 \quad $\dashrightarrow$ \quad
 &
 \AXC{$\seq{\Gamma}{\phi}$}
 \LeftLabel{$w$}
 \UIC{$\seq{\phi\to\chi,\Gamma,\Delta}{\phi}$}
 \AXC{$\seq{\chi,\Delta}{\psi}$}
 \LeftLabel{$w$}
 \UIC{$\seq{\chi,\Gamma,\Delta}{\psi}$}
 \LeftLabel{$\to_l$}
 \BIC{$\seq{\phi\to\chi,\Gamma,\Delta}{\psi}$}
 \DP
\end{tabular}
\end{adjustbox}
\end{center}
In this way, we obtain a derivation in $\Gmc\nfour$ with (possibly) applications of weakening and contraction. According to Proposition \ref{prop:weakening-contraction-admissibility}, weakening and contraction are admissible in $\Gmc\nfour$, thus we can further eliminate these applications and in the end, obtain one derivation in $\Gmc\nfour$.

Next, we finish the proof of cut-admissibility of $\cnfckone$.
It remains to consider the first, third, and fourth cases.

If $\phi\in\{p,\simm p\}$ for some $p\in\Prop$ and $\phi\in\Gamma$, we obtain the conclusion $\seq{\Gamma,\Delta}{\chi}$ from the right premise $\seq{\phi,\Delta}{\chi}$ by weakening. If $\phi\in\{p,\simm p\}$ for some $p\in\Prop$ and $\phi=\chi$, we obtain the conclusion $\seq{\Gamma,\Delta}{\chi}$ from the left premise $\seq{\Gamma}{\phi}$ by weakening. Otherwise, $\seq{\Delta}{\chi}$ is an axiom, we apply weakening to it and obtain the conclusion as required.

In the \textbf{third case}, $\phi$ is principal in the left premise \textit{only}. If $\phi$ is \textit{weak} in the right premise, which means $\Rmc_2$ is one of the modal rules and is neither $\Box$-prefixed nor $\simm\lozenge$-prefixed. We obtain the conclusion of $\cut$ from the right premise directly. For example, if $\Rmc_2=\Boxjoin$ and $\chi=\Box\theta$, we transform the derivation as follows:
\begin{align*}
\vlderivation{
\vliin{}{\cut}{\seq{\Gamma,\Delta}{\Box\theta}}{
\vltr{\Dmc_1} {
\seq{\Gamma}{\phi}
 }{
 \vlhy {\quad}}
 {
 \vlhy {}}
 {
 \vlhy {\quad}}
 }
 {
 \vlin{\Boxjoin}{}{\seq{\phi,\Delta}{\Box\theta}}{
 \vltr {\Dmc_2'} {
 \seq{\Delta^\Box,\diasimm{\Delta}}{\theta}
 }{
 \vlhy {\quad}}
 {
 \vlhy {}}
 {
 \vlhy {\quad}} 
 }
 }
 }
 &
\dashrightarrow
\vlderivation{
 \vlin{w}{}{\seq{\Gamma,\Delta}{\Box\theta}}{
 \vlin{\Boxjoin}{}{\seq{\Delta}{\Box\theta}}{
 \vltr {\Dmc_2'} {
 \seq{\Delta^\Box,\diasimm{\Delta}}{\theta}
 }{
 \vlhy {\quad}}
 {
 \vlhy {}}
 {
 \vlhy {\quad}} 
 }
 }
 }
\end{align*}
If $\phi$ is \emph{side} in the right premise, note that there are no side formulas in a~mo\-dal rule, so $\Rmc_2$ is a rule in $\Gmc\nfour$. By the invertibility of $\Gmc\nfour$ rules, we unfold $\Dmc$ and transform it as follows:
\begin{align*}
\vlderivation{
 \vliin{}{\cut}{\seq{\Gamma,\Delta}{\chi}}{
 \vltr {\Dmc_1} {
 \seq{\Gamma}{\phi}
 }{
 \vlhy {\quad}}
 {
 \vlhy {}}
 {
 \vlhy {\quad}}
 }
 {
 \vlin{\Rmc_2}{}{\seq{\phi,\Delta}{\chi}}{
 \vltr {\Dmc_2'} {
 \seq{\phi,\Delta'}{\chi'}
 }{
 \vlhy {\quad}}
 {
 \vlhy {}}
 {
 \vlhy {\quad}} 
 }
 }
 }
&\dashrightarrow
\vlderivation{
 \vlin{\Rmc_2}{}{\seq{\Gamma,\Delta}{\chi}}{
 \vliin{}{\cut}{\seq{\Gamma,\Delta'}{\chi'}}{
 \vltr {\Dmc_1} {
 \seq{\Gamma}{\phi}
 }{
 \vlhy {\quad}}
 {
 \vlhy {}}
 {
 \vlhy {\quad}}
 }
 {
 \vltr {\Dmc_2'} {
 \seq{\phi,\Delta'}{\chi'}
 }{
 \vlhy {\quad}}
 {
 \vlhy {}}
 {
 \vlhy {\quad}} 
 }
 }
 }
\end{align*}

Finally, in the \textbf{fourth case}, $\phi$ is \emph{not principal} in the left premise. Since sequents have only one formula on the right, $\phi$ can only occur as \textit{side} formula in the application of $\Rmc_1$, which implies $\Rmc_1$ must be a left rule in the basic calculus $\Gmc\nfour$ (for $\to_l$ only occur as side formula in the right branch). We have the following general form of $\Dmc$ that we transform 
as shown below (weakening is omitted):
\begin{align*}
            \vlderivation{
            \vliin{}{\cut}{\seq{\Gamma,\Delta}{\chi}}{
            \vlin{\Rmc_1}{}{\seq{\Gamma}{\phi}}{
            \vltr {\Dmc_1'} {
                    \seq{\Gamma'}{\phi}
                }{
                \vlhy {\quad}}
                {
                \vlhy {}}
                {
                \vlhy {\quad}}
            }
            }
            {
                \vltr {\Dmc_2} {
                    \seq{\phi,\Delta}{\chi}
                }{
                \vlhy {\quad}}
                {
                \vlhy {}}
                {
                \vlhy {\quad}} 
            }
            }
&\dashrightarrow
\vlderivation{
 \vlin{\Rmc_1}{}{\seq{\Gamma,\Delta}{\chi}}{
 \vliin{}{\cut}{\seq{\Gamma',\Delta}{\chi}}{
 \vltr {\Dmc_1'} {
 \seq{\Gamma'}{\phi}
 }{
 \vlhy {\quad}}
 {
 \vlhy {}}
 {
 \vlhy {\quad}}
 }
 {
 \vltr {\Dmc_2} {
 \seq{\phi,\Delta}{\chi}
 }{
 \vlhy {\quad}}
 {
 \vlhy {}}
 {
 \vlhy {\quad}} 
 }
 }
 }
\end{align*}
This completes the proof. \qed
\end{proof}
\sequentcompleteness*
\begin{proof}
The completeness is obtained as follows. $\Hmc\Lmsf$-axioms can be proved in the corresponding $\Gmc\Lmsf$  (cf.~Section~\ref{sec:sequents} for examples). The modal rules can be derived by corresponding rules in $\Gmc\Lmsf$ with the invertibility of $\to_r$. Finally, $\mathbf{mp}$ can be simulated via the $\cut$ rule which is admissible by Theorem~\ref{theorem:cutadmissibility}.
\qed
\end{proof}